\documentclass[11pt,twoside]{article}
\usepackage{newpasp}

\usepackage{epsf}

\index{Mihos, C.}

\begin{document}

\title{Interactions and mergers at higher redshift}
\author{Chris Mihos}
\affil{Case Western Reserve University, Cleveland, OH 44106}


\begin{abstract}

In models of hierarchical structure formation, interactions and
mergers at high redshift play a key role in the formation and
evolution of galaxies. Numerical modeling and observations of nearby
systems explore the detailed physics of galaxy interactions, but
applying these results to the high redshift universe remains
complicated. Evolution in the properties of galaxies over cosmic
history may lead to differences in interaction-induced star formation
as a function of redshift. High redshift interacting galaxies may be more
prone to extreme, disk-wide starburst activity than are low redshift
systems. Also, while activity in nearby clusters seems to be largely
diminished, at higher redshift we observe galaxies falling into
clusters for the first time. A variety of processes, including
interactions and mergers, can drive strong evolution in cluster
galaxies at higher redshift.

\end{abstract}




\section{Lessons from the Nearby Universe}

Most of our detailed knowledge of the effects interactions have on
galaxies comes from studies grounded in the local
universe. Well-studied samples of interacting galaxies such as the
Toomre (1977) sequence extend only out to $\sim 10,000$ km/s, and so
much spatial information is lost at redshifts greater than a few
tenths that we can only rely on statistical studies of interactions at
higher redshift. Dynamical models, which have illuminated many of
the physical processes involved during galaxy collisions, rely heavily
on galaxy models constructed to resemble nearby galaxies in terms of
their structural and kinematic properties.  Whether such models
accurately portray interacting galaxies at high redshift is unclear --
structural differences between high and low redshift galaxies may
translate to substantial differences in the response to an
interaction.

Nonetheless, these dynamical models allow us to view the basic
evolution of a merging encounter (Figure 1, from Mihos \& Hernquist
1996 [MH96]). Shortly after the initial
passage, the galaxies become strongly distorted, with disk self-gravity
amplifying the perturbation into strong spiral arms or central
bars. As the galaxies separate, dynamical friction between the dark
halos slows the galaxies on their orbit, causing them to fall back and
merge together.  The violent relaxation accompanying the merger
destroys the disks and results in the creation of an elliptical-like
remnant. During the interaction and ensuing merger, a large fraction
of the galaxies' ISM is compressed and driven inwards due largely to
gravitational torques during the encounter. This compression and
inflow presumably leads to strong starburst and/or AGN activity during the 
encounter.

\begin{figure}
\plottwo{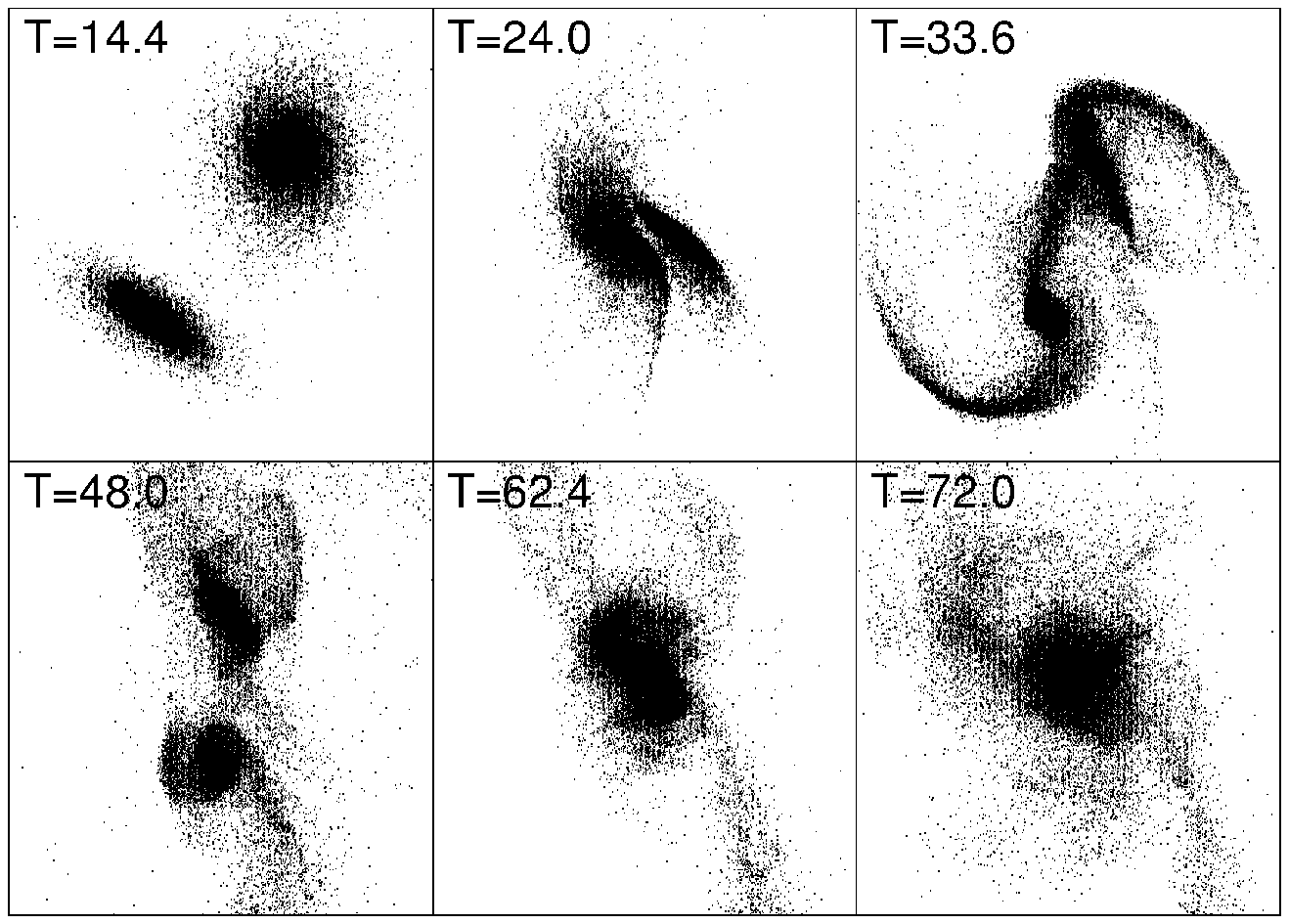}{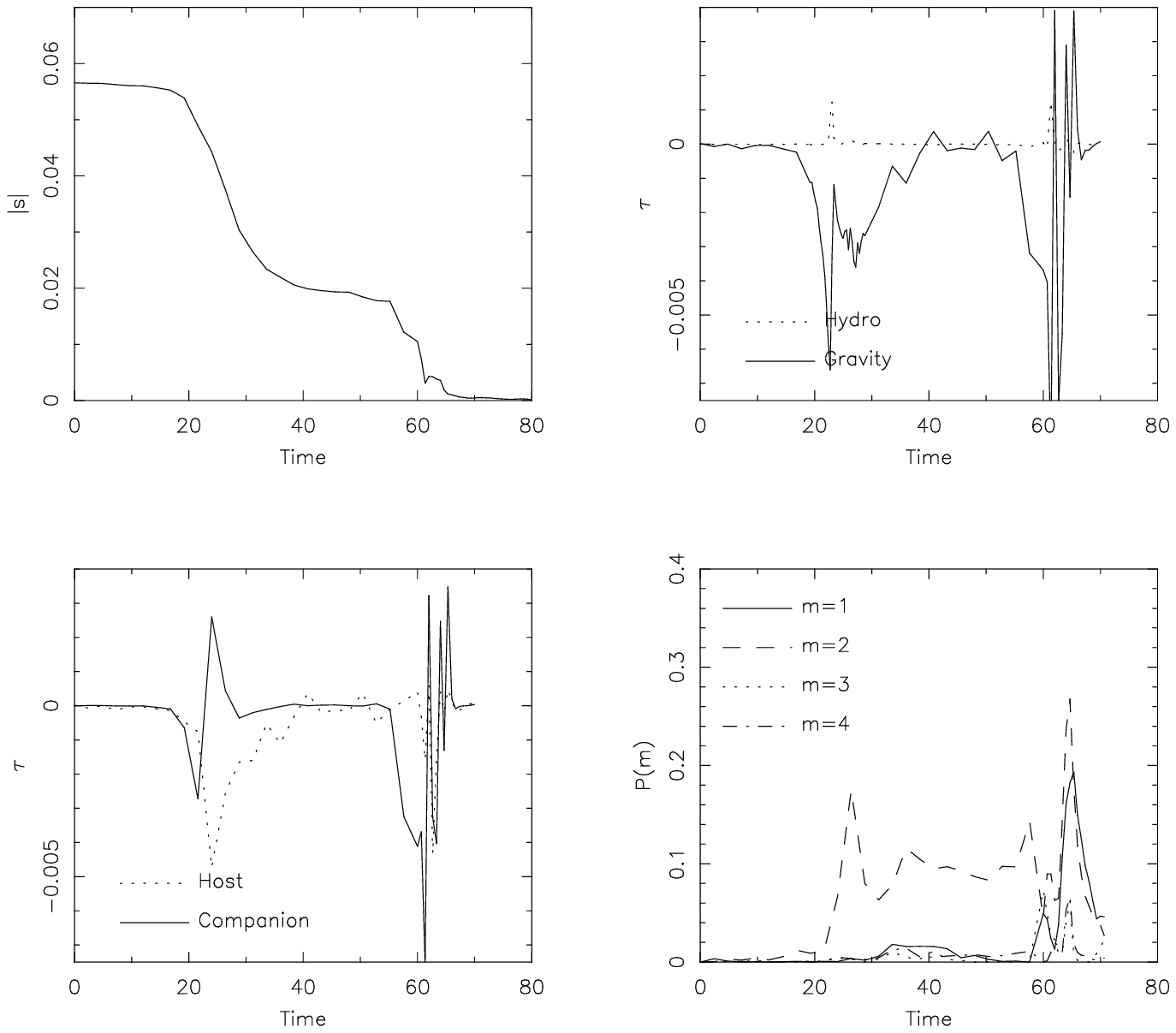}
\caption{Left: Evolution of an equal mass merger, from MH96. Time is given
in model units; the sequence covers approximately 750 Myr. Right:
Evolution of the inflowing gas. The subpanels show (top left) the
angular momentum of the gas, (top right) the gravitational and
hydrodynamic torques acting on the gas, (bottom left) the gravitational
torques from the host and companion galaxies respectively, and (bottom
right) the growth of nonaxisymmetric modes in the stellar disk.}
\label{fig1}
\end{figure}

The triggering of enhanced star formation by interactions is borne out
in the observational record, but the effects, on average, are not
extreme. Optically selected samples of interacting galaxies show star
formation rates elevated only by factors of a few over normal
galaxies, with inferred starburst mass fractions of only a few percent
(e.g., Larson \& Tinsley 1978; Kennicutt et al. 1987).  This enhanced
SFR is preferentially found in the nuclear regions of galaxies (Keel
et al. 1985; Kennicutt et al. 1987), suggesting that nuclear inflow is the
preferred method for triggering starbursts in interacting systems at
low redshift.  The most extreme examples of merger-induced activity
are found in infrared samples; indeed, the most luminous starbursts in
the local universe, ultraluminous infrared galaxies (ULIRGs) with luminosities
$> 10^{12}$ L$_{\sun}$ all show evidence of being interacting systems
in the final throes of merging (Sanders \& Mirabel 1996). Yet it is
important to realize that not all merging systems are ultraluminous --
the link between mergers and luminous activity is not a simple one.

Given the wide range of star-forming properties exhibited by
interacting systems, what determines how a galaxy responds to
an interaction? Dynamical modeling has shown that stability of the
host disk plays an essential role in the degree of star-forming
response. Also
shown in Figure 1 is the evolution of the gas in the prograde disk
during the encounter.  Two distinct phases of inflow are seen -- one
early, and a second late in the encounter when the galaxies ultimately
merge.  By decomposing the torques which drive this gas inwards, we
see that they are largely gravitational in origin, and that the first
phase of inflow is driven by gravitational instabilities in the host
disk rather than from the passing companion. The rapid inflow shown in
Figure 1 is the result of bars and strong spiral features in the host
disk; in these features, the gas tends to slightly lead the stellar
component (Barnes \& Hernquist 1996 [BH96]), resulting in a strong
gravitational torque which drives the gas inwards. In this model, the
first phase of inflow is halted by the presence of a central bulge
which stabilizes the inner disk; when the galaxies then ultimately
merge this stability is overwhelmed by the merger process and a second
phase of inflow commences. In models without a bulge, instabilities
are so strong that the gas is driven to the nucleus in a single
phase at early times, while the galaxies are still well-separated.

Because nuclear gas inflows are so intimately tied to instabilities in
the galactic disk, star formation in interacting galaxies should
depend strongly on the stability of disks against strong
perturbations.  This idea has been demonstrated in a number of
theoretical studies. A variety of mechanisms have been shown to
suppress instabilities and slow the inflow, including the presence of
a dense central bulge (MH96), a lowered disk surface density relative
to the dark matter content (Mihos et al. 1997), and retrograde encounter
geometries, which reduce the spin-orbit coupling driving the
instability (BH96). In these cases, early inflow is suppressed in
favor of milder gas compressions (and, presumably, milder star
formation enhancement) throughout the disk.

 

\section{Going to Higher Redshift}

Given that the response of a galaxy to an interaction is such a
strong function of its internal structure, the question
of how high redshift interactions behave becomes twofold:
\begin{itemize}
\item How do the properties of high redshift galaxies compare to those at
$z=0$?
\item How do these differences translate to differences in the collisional
response?
\end{itemize}

If the structural and kinematic properties of high redshift galaxies
were known, we could use simple analytic perturbation theory to begin
to probe their response to interaction. Disk stability depends on a
variety of properties, as illustrated by the Toomre $Q$ and $X_2$
parameters for local and global (bar) instabilities:
$$Q = { {\sigma_r \kappa} \over {3.36 G \Sigma} } \hskip 1.0truein X_2
= { {\kappa^2 R} \over {4\pi G \Sigma} }$$ where $\sigma_r$ is the
radial velocity dispersion, $\kappa$ is the epicyclic frequency,
$\Sigma$ is the disk surface density, and $R$ is disk
radius. Stability is ensured in both cases if $Q>1, X_2>1$\footnote{In
the case of $X_2$, this holds for a linearly rising rotation curve. If
the rotation curve is flat, the criteria becomes $X_2>3$.} (Toomre
1964, 1981). While the linear perturbation theory behind these
criteria is surely inadequate to describe the self-gravitating
response to a strong encounter, it at least provides a qualitative
guide for galaxy behavior: disks with a rapidly rising
rotation curve (due to a kinematically hot component such as a bulge)
and/or a lower surface density should be preferentially more stable
systems.
 
Unfortunately, the structural and dynamical properties of high
redshift galaxies are still  poorly constrained.  One reasonable
presumption is that high redshift galaxies are more gas rich.  For
example, in an $\Omega_M=0.3, \Omega_\Lambda=0.7$ cosmology, a spiral
galaxy formed at $z_f=3$ with an exponentially decaying star formation
rate with decay timescale $\tau=5$ Gyr has a gas fraction of $f_g=0.1$
at $z=0$ and $f_g=0.5$ at $z=1$. This heightened gas fraction may 
lead to a stronger disk-wide response to an 
interaction. Semianalytic models (e.g., Kauffmann 1996) have suggested
that in addition to being gas rich, high redshift galaxies may lie
closer to a threshold density -- related to the Toomre $Q$ parameter
-- for triggering star formation (see Quirk 1972; Kennicutt 1989).
If high redshift galaxies are gas-rich and prone to local stabilities, 
interactions may drive disk gas to collapse locally, igniting intense
star formation throughout the disk rather than preferentially driving
nuclear starbursts.

Indeed, we may be seeing such an effect in the very knotty appearance
of high redshift galaxies. The fraction of peculiar and/or interacting
system rises significantly at higher redshift (e.g., Glazebrook et al.
1995; Abraham et al. 1996; Ferguson et al. 2000), and these
peculiarities are often manifested as multiple high surface brightness
knots distributed throughout a common envelope. With large gas mass
fractions, high redshift galaxies can respond quite strongly to even
relatively mild interactions which, at low redshift, incite a much
more subdued response.

What about instability-driven inflows and {\it nuclear} starbursts -- are
high-z disks any more or less stable than low-z disks? Unfortunately,
a robust answer to this question requires detailed knowledge of the
kinematics and structural properties of high-z disks, information
which is not yet available. However, a number of clues hint at
answers.  First, out to redshifts of $z\sim 1$ there is evidence for
only mild evolution in the structural properties of luminous disk
galaxies (Simard et al. 1999; Vogt 2000).  At higher redshifts
($z>1$), there is some observational evidence for a decrease in the
bulge-to-disk ratio of galaxies (see, e.g., the compilation in Marleau
\& Simard 1998), although these studies have extremely complicated
selection effects (Giavalisco et al. 1996). Nonetheless these studies
are consistent with theoretical models which suggest that the bulges
of luminous spirals are largely in place by a redshift of $z\sim 1$
(Kauffmann 1996; Baugh et al. 1996). At the highest redshift, little
structural information can be gleaned, but evidence does exist that
the galaxy populations are more compact than at low redshift (Lowenthal
et al. 1997; Giavalisco et al. 1996).

So where does this leave us? Based on these arguments, a cartoon picture 
of interaction induced star formation might look something like this:

\begin{itemize}
\item {\bf $z > 1-2$:} At high redshift, galaxies are likely to be very
unstable. Their large gas fractions leave them prone to violent, local 
instabilities, and if bulges are not yet in place, global instabilities
as well. Even mild interactions may drive intense starbursts involving a
substantial fraction of the gas content of galaxies, and interactions
may be responsible for some of the most luminous  SCUBA sources
(Blain et al. 1999; Ivison et al. 2000).

\item {\bf $0.5 < z < 1$:} As disks build up their bulges, they begin
to stabilize against violent, global instabilities. Yet their heightened
gas content leaves them still subject to fairly strong disk starbursts,
as evidenced by their knotty appearance. Nonetheless, the absolute star
formation rates are not extreme; studies of galaxies at moderate redshifts 
indicate typical SFR enhancements of a factor of a few (Le Fevre et al. 2000),
similar to that observed locally (Kennicutt et al. 1987).

\item {\bf $z < 0.5$:} At lower redshift, as gas reservoirs are
depleted, galaxies evolve toward more quiescent interactions. This is
reflected in the modest SFR enhancements of low redshift interactions
(Kennicutt et al. 1987), although the ULIRG samples demonstrate that if
the encounter leads to a merger, intense starbursts ensue (Sanders \&
Mirabel 1996).

\end{itemize}

Having made this cartoon scenario, it is now time to rip it apart --
unfortunately, the situation is nowhere near as simple as spelled out
above. Stability is not simply a function of bulge-to-disk ratio, but
depends rather on the mass distribution and kinematics of galaxies.
Compare for example the models of MH96 with those of BH96, both of
which used galaxies with B/D=1/3, but with differing bulge
densities. In the MH96 models, the inflow response was rather
insensitive to orbital geometry, as the presence or absence of a dense
bulge dominated the dynamics of the inner few kpc of the galaxies. In
contrast, the more diffuse bulges of BH96 provided comparatively less
stability, so that the disks were more vulnerable to interaction driven
instabilities. This uncertainty is reflected in the fact that there is a
very large scatter in star forming properties along the Hubble
sequence and that star formation appears more closely tied to global gas
content than bulge-to-disk ratio (Kennicutt 1998). Indeed the
crucial information needed to understand the stability properties of high
redshift disks is high quality, spatially resolved kinematic and
structural data on galaxies as a function of redshift.

Other questions abound as well. How are bulge and disk formation
linked?  If bulges form first, and disks grow around them through gas
accretion into the potential well, the disks should begin life
globally stable, and move towards instability as the disk surface
density increases. Also, what role does secular evolution play in the
formation of bulges?  Models of disk galaxy evolution have suggested
that some bulges form as a result of instabilities in the stellar disk
(e.g., Pfenniger \& Norman 1990; Norman et al. 1996; Corteau et
al. 1996). Under these models, disks form first, while bulges build up continuously over
cosmic time, so that the disk galaxy population as a whole may gradually
become {\it more} stable against strong inflow and starbursts. Unfortunately, the
stabilizing properties of these ``secular'' bulges is unclear ---
growing from material originally in the disk, these bulges may have a
faster rotation than true r$^{1/4}$ bulges (Kormendy 1993) and may
not provide strong dynamical stability.

\section{Cluster Environments}

The fact that our knowledge of interactions comes largely from the
local universe also means that we are largely observing interactions
between field galaxies, or galaxies within evolved clusters. When
looking at galaxy clusters at higher redshift, we see clusters which
are dynamically young, and galaxies which may be falling into clusters
for the first time. A variety of dynamical mechanisms operate on
galaxies within clusters:
{\it cluster tides} act to strip the outer regions of cluster
galaxies (e.g., Malamuth \& Richstone 1984), and drive instabilities
in infalling disks (Byrd \& Valtonen 1990);
{\it collisions and mergers} of galaxies in the high density
environment drive evolution in the galaxy populations;
{\it high speed encounters} between galaxies in clusters 
impulsively heat galaxies (e.g., Moore et al. 1996); and
{\it ram pressure stripping} may compress and remove the ISM
from infalling disks, first igniting a starburst, then rapidly
shutting it down (Gunn \& Gott 1972).

\begin{figure}
\plotone{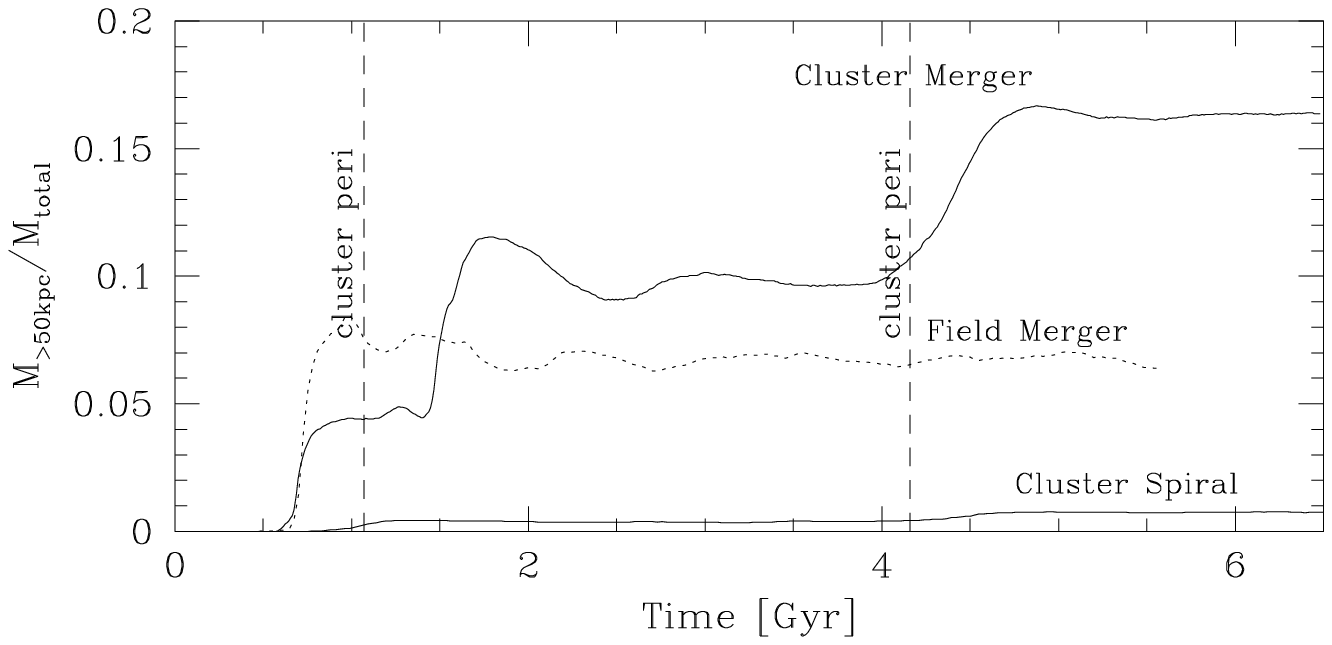}
\caption{The evolution of tidally stripped material (defined as the
fraction of disk material at radius $r>$50 pc) in different
environments. The ``field merger''
curve shows the fraction of material lost from a major merger of two
spiral galaxies in
the field environment (see Figure 1). The ``cluster merger'' curve
shows the same merger, but this time occurring in a Coma-like cluster 
potential (Geller et al. 1999). In this potential, the galaxy pair orbits
with $r_{peri}=0.5$ Mpc and 
$r_{apo}=2.0$ Mpc. The ``cluster spiral'' curve shows stripping from
a single non-interacting spiral galaxy on the same cluster orbit.}
\label{fig1}
\end{figure}

Before addressing how cluster interactions evolve in detail, we should
ask, given the high velocity dispersions of clusters, so
we expect any slow collisions or mergers at all?  Certainly an
individual galaxy traveling at high speed through the cluster
potential will have little chance of a slow interaction with another
cluster member. However, clusters grow hierarchically, by accreting
small groups of galaxies, and interactions within the infalling groups
may occur before the group is disrupted by the tidal field of the
cluster. N-body simulations of the formation of rich clusters shows
clearly that this process occurs (e.g., Dubinski 1998), and we see
evidence for slow interactions in clusters at moderate redshift (Couch
et al.  1998; van Dokkum et al. 2000). The rate of accretion onto
clusters is a strong function of redshift and cosmology; simulations
by Gighna et al. (1998) suggest that mergers within the cluster virial
radius are largely shut off at redshifts $z < 0.5$. Unlike studies of
nearby clusters, at higher redshifts slow interactions may play a very
important role in the evolution of cluster galaxies.

Interactions and mergers in the cluster environment should differ from
field collisions in a number of ways. Of particular interest is the
role of the cluster tides and hot ICM on the diffuse tidal material
generated during a low speed collision. In the field, the overwhelming
majority of the tidally liberated gas and stars in mergers remains
bound to the remnant, falling back over several Gyr, and leaving a
long-lived record of the interaction on the remnant (Hibbard \& Mihos
1995). The continued reaccretion of gas by the remnant may result in
the HI disks observed in several nearby field ellipticals, and perhaps
may even result in the delayed formation of S0s (e.g., Schweizer
1998). In contrast, the cluster environment will rapidly remove this
material from any merger remnant in the cluster, feeding both the hot
ICM and the diffuse intracluster starlight (see Figure 2). In contrast
to traditional picture of cluster tidal and ram pressure stripping --
which affect primarily the outskirts of galaxies -- cluster
interactions and mergers act to accentuate the process by ``dredging
up'' material from the inner regions of galaxies and moving it to large
radius, where the cluster tides and (in the case of tidally liberated
HI) hot gas can efficiently strip it. In the context of S0 formation
models, delayed formation of S0s through reaccretion of tidal gas
seems not to be a viable option in the cluster environment. However,
the disk heating associated with galaxy interactions (e.g., Walker
et al. 1996) could still lead
to the formation of disky S0s in clusters.

Finally, the role of ram pressure stripping (RPS) in driving galaxy
evolution remains unclear. Several authors have invoked RPS as a way
of truncating star formation and forming E+A and S0 galaxies in
clusters. However, RPS affects predominantly the outer HI disks of
galaxies, and should leave the inner molecular gas intact. Indeed,
recent RPS simulations by Quilis et al. (2000) only succeeded in
removing the ISM of cluster galaxies in models where the gas
distribution in the disk had a central hole; these authors posit that
the central molecular gas would rapidly dissociate due to a triggered
starburst, and be swept from the disk over very short timescales
($\Delta t \sim 10^7$ years).  However, from an observational
standpoint, this appears not to be the case. The molecular gas in
cluster galaxies seems to survive the cluster environment: Kenney \&
Young (1989) find that HI deficient Virgo spirals contain significant
quantities of molecular gas, while Casoli et al. (1998) find no CO
deficiency in cluster spirals.  Interestingly, there is evidence
that some poststarburst galaxies in clusters may still have appreciable
obscured star formation (Smail et al. 1999), arguing that star formation
in cluster spirals is extended in duration ($\Delta t \sim$ several
$\times\ 10^8 - 10^9$ years).  Given the fact that cluster interactions
and mergers occur, and that cluster galaxies can retain some fraction
of their gas for an appreciable time, it seems premature to abandon
collisional formation mechanisms for the formation of some cluster E+A and
S0 galaxies at higher redshift.

This research is sponsored in part by an NSF Career Fellowship and 
the San Diego Supercomputing Center.


\begin{references}
{\small
\refpar Abraham, R.G. et al. 1996, MNRAS, 279, L47
\refpar Barnes, J.E. \& Hernquist, L. 1996, ApJ, 471, 115 (BH96)
\refpar Baugh, C.M., Cole, S., \& Frenk, C.S. 1996, MNRAS, 283, 1361
\refpar Blain, A.W. et al. 1999, MNRAS, 309, 715
\refpar Byrd, G. \& Valtonen, M.\ 1990, ApJ, 350, 89
\refpar Casoli, F. et al. 1998, A\&AS, 331, 451 
\refpar Corteau, S., de Jong, R.S., \& Broeils, A.H. 1996, ApJ, 457, L73
\refpar Couch, W.J. et al. 1998, ApJ, 497, 188
\refpar Dubinski, J. 1998, ApJ, 502, 141
\refpar Ferguson, H.C., Dickison, M., \& Williams, R. 2000, astro-ph/0004319
\refpar Geller, M.J., Diaferio, A., \& Kurtz, M.J. 1999, ApJ, 517, L23
\refpar Giavalisco, M. et al. 1996, AJ, 112, 369
\refpar Giavalisco, M., Steidel, C.C., \& Machetto, F.D. 1996, ApJ, 470, 189
\refpar Ghigna, S. et al. 1998, MNRAS, 300, 146 
\refpar Glazebrook, K., Ellis, R., Santiago, B., \& Griffiths, R. 1995, 
MNRAS, 275, L19
\refpar Gunn, J.E., \& Gott, J.R. 1972, ApJ, 176, 1
\refpar Hibbard, J.E. \& Mihos, J.C. 1995, AJ, 110, 140
\refpar Ivison, R.J. et al. 2000, MNRAS, 315, 209
\refpar Kauffmann, G. 1996, MNRAS, 281, 487
\refpar Keel, W.C., Kennicutt, R.C., Hummel, E., \& van der Hulst, J.M.
1985, AJ, 90, 708
\refpar Kenney, J.D.P., \& Young, J.S. 1989, ApJ, 344, 171
\refpar Kennicutt, R.C. 1989, ApJ, 344, 685
\refpar Kennicutt, R.C. 1998, ARAA, 36, 189
\refpar Kennicutt, R.C. et al. 1987, AJ, 93, 1011
\refpar Kormendy, J. 1993, in IAU Symp. 153: Galactic Bulges, eds. H.Dejonghe
\& H.J. Habing (Dordrecht: Kluwer), 209
\refpar Larson, R.B., \& Tinsley, B.M. 1978, ApJ, 219, 46
\refpar Le Fevre, O. et al. 2000, MNRAS, 311, 565
\refpar Malumuth, E M. \& Richstone, D.O. 1984, ApJ, 276, 413 
\refpar Marleau, F.R., \& Simard, L. 1998, ApJ, 507, 585
\refpar Mihos, J.C., \& Hernquist, L. 1996, ApJ, 464, 641 (MH96)
\refpar Mihos, J.C., McGaugh, S.S., \& de Blok, W.J.G. 1997, ApJ, 477, L79
\refpar Moore, B. et al. 1996, Nature, 379, 613
\refpar Norman, C.A., Sellwood, J.A., \& Hasan, H. 1996, ApJ, 462, 114
\refpar Pfenniger, D., \& Norman, C.A. 1990, ApJ, 363, 391
\refpar Quilis, V., Moore, B., \& Bower, R. 2000, Science, 288, 1617
\refpar Quirk, W.J., 1972, ApJ, 176, L9
\refpar Sanders, D.B., \& Mirabel, I.F. 1996, ARAA, 34, 749
\refpar Schweizer, F. 1998, in {\it Galaxies: Interactions and Induced 
Star Formation}, (Berlin: Springer), 105
\refpar Simard, L. et al. 1999, ApJ, 519, 563
\refpar Smail, I. et al. 1999, ApJ, 525, 609
\refpar Toomre, A. 1964, ApJ, 139, 1217
\refpar Toomre, A. 1977, in The Evolution of Galaxies and Stellar Populations,
ed. B. Tinsley \& R. Larson (New Haven: Yale Univ. Press), 401
\refpar Toomre, A. 1981, in The Structure and Evolution of Normal
Galaxies, ed. S.M. Fall \& D. Lynden-Bell (London: Cambridge Univ. Press), 111
\refpar van Dokkum, P. et al. 2000, ApJ, 541, 95
\refpar Vogt, N.P. 2000, in ASP Conf. Ser. TBD, Gas and Galaxy Evolution, 
ed. J.E. Hibbard, M.P. Rupen, \& J.H. van Gorkom (San Francisco: ASP)
\refpar Walker, I.R., Mihos, J.C., \& Hernquist, L. 1996, ApJ, 460, 121 

}
\end{references}
\end{document}